\documentclass[reprint,showpacs,preprintnumbers,amsmath,amssymb,amsfonts]{revtex4-2}
\usepackage{graphicx,latexsym,amssymb,amsmath,color,multirow,mathrsfs,booktabs,ifpdf}
\def\bm#1{\mbox{\boldmath $#1$}}
\begin{document}
\title{Spin symmetry energy and equation of state of the spin-polarized neutron star matter}
\author{Nguyen Hoang Dang Khoa$^1$}
\author{Ngo Hai Tan$^{2,3}$}
\author{Dao T. Khoa$^4$}
\affiliation{$^1$ University of Science and Technology of Hanoi, 
  Hanoi 100000, Vietnam. \\
$^2$ Faculty of Fundamental Sciences, Phenikaa University, Hanoi 12116, Vietnam.\\
$^3$Phenikaa Institute for Advanced Study (PIAS), \\ 
  Phenikaa University, Hanoi 12116, Vietnam. \\
$^4$Institute for Nuclear Science and Technology, VINATOM,
  Hanoi 100000, Vietnam.}
\date{\today}
\begin{abstract}
Equation of states (EOS) of the spin-polarized nuclear matter (NM) is studied 
within the Hartree-Fock (HF) formalism using the realistic density dependent 
nucleon-nucleon interaction. With a nonzero fraction $\Delta$ of spin-polarized 
baryons in NM, the spin- and spin-isospin dependent parts of the HF energy
density give rise to the \emph{spin symmetry} energy that behaves in about 
the same manner as the \emph{isospin symmetry} energy, widely discussed 
in literature as the nuclear symmetry energy. The present HF study shows 
a strong correlation between the spin symmetry energy and nuclear symmetry energy 
over the whole range of baryon densities. The important contribution of the spin 
symmetry energy to the EOS of the spin-polarized NM is found to be comparable 
with that of the nuclear symmetry energy to the EOS of the isospin-polarized or
asymmetric (neutron-rich) NM. Based on the HF energy density, the EOS of the 
spin-polarized ($\beta$-stable) np$e\mu$ matter is obtained for the determination 
of the macroscopic properties of neutron star (NS). A realistic density dependence 
of the spin-polarized fraction $\Delta$ have been suggested to explore the impact 
of the spin symmetry energy to the gravitational mass $M$ and radius $R$, as well 
as the tidal deformability of NS. Given the empirical constrains inferred from 
a coherent Bayesian analysis of gravitational wave signals of the NS merger 
GW170817 and the observed masses of the heaviest pulsars, the strong impacts 
of the spin symmetry energy $W$, nuclear symmetry energy $S$, and nuclear 
incompressibility $K$ to the EOS of nucleonic matter in magnetar were revealed. 
\end{abstract}
\pacs{}
\maketitle
\section{Introduction}\label{sec1} 
The rotating neutron stars are known to possess strong magnetic field 
\cite{Lat07,Brod00,Dex17}, with the field strength $B$ of the order of $10^{14}$ 
to $10^{19}$ G, so that the effects of magnetic field on the equation of states (EOS) 
of NS matter should not be negligible. In particular, a significant fraction of baryons 
in NS matter might have their spins polarized along the axis of magnetic field. The 
full spin polarization of neutrons was shown by Broderick {\it et al.} \cite{Brod00}
to likely occur at the high field strength of $B\gtrsim 4.41\times 10^{18}$ G. It is 
commonly assumed that the magnetic field of NS is usually much weaker than the upper
limit of $B\approx 10^{19}$ G, and the spin polarization of baryons is often 
neglected in the mean-field studies of the EOS of NS matter. Recently, the ``blue''
kilonova ejecta observed in the aftermath of the NS merger GW170817 \cite{Abb17a,Abb17b,Evan17} 
have been suggested by Metzger {\it et al.} \cite{Metzger18} to be caused by both the 
$\gamma$ decay of the \textit{r}-process nuclei and magnetically accelerated wind from 
the strongly magnetized hypermassive NS remnant. A rapidly rotating hypermassive NS remnant 
with the magnetic field of $B\approx (1–3)\times 10^{14}$ G at the surface has been assumed  
to explain the velocity, total mass, and enhanced electron fraction of the kilonova 
ejecta \cite{Metzger18}. Such a scenario seems to agree with the prediction made by
Fujisawa {\it et al.} \cite{Fuji14} for the strength of magnetic field in the outer core 
of magnetar, and a partial spin polarization of baryons might well occur in the two 
merging neutron stars of GW170817. To investigate such effects, a nonrelativistic 
Hartree-Fock (HF) study of the spin-polarized nuclear matter (NM) has been done 
recently \cite{Tan20}, assuming different (relative) strengths $\Delta$ of the spin 
polarization of baryons. The EOS obtained in the HF approach for NS matter consisting 
of strongly interacting baryons and leptons, i.e., the np$e\mu$ matter in 
$\beta$-equilibrium was used as input to determine the gravitational mass $M$ and 
radius $R$ of NS \cite{Tan20,Tan21} from the solutions of the Tolman-Oppenheimer-Volkoff 
(TOV) equations \cite{Oppenheimer39}.   

Given a realistic EOS of NS matter, General Relativity not only explains the compact 
shape of NS in the hydrostatic equilibrium but also predicts interesting behaviors of NS 
in the strong gravitational field formed by two inspiraling neutron stars during their
merger. In particular, the shape of each NS is tidally deformed by the mutual attraction 
of two coalescing neutron stars to gain nonzero multipole moments \cite{Hind08,Hind10,Damour09}, with the energy being lost via the emission of gravitational waves (GW). The tidal 
deformation is usually expressed in terms of the tidal Love number $k_2$ of the second order. 
which has been inferred recently from the analysis of the observed GW signals from GW170817, 
and translated into the constraint for the gravitational mass $M$ and radius $R$ of NS 
\cite{Abb18}. Because this empirical constraint serves now as an important reference 
in validating different models of the EOS of NS matter, we have applied in the present work 
the HF approach suggested in Ref.~\cite{Tan20} to study in more detail the impact by the spin 
polarization of baryons to the EOS of NS matter. Governed by the same SU(2) symmetry, the
spin symmetry energy $W$ is shown in Sect.~\ref{sec2} to behave in about the same manner 
as the isospin symmetry energy $S$, widely known as the \emph{nuclear symmetry energy}. 
In particular, the parabolic approximation is valid also for the spin symmetry energy, 
so that the (repulsive) contribution from $W$ to the total NM energy is directly proportional 
to $\Delta^2$. An interesting correlation between the slope parameter $L$ of the nuclear 
symmetry energy $S$ and the slope $L_{\rm s}$ of the spin symmetry energy $W$ has been 
found and discussed in detail. 

The EOS of the $\beta$-stable spin polarized NS matter obtained in Sect.~\ref{sec3}
is used as the input to solve the linearized Einstein equation for the metric perturbation 
of the stress-energy tensor of NS to determine the tidal deformability 
$\Lambda$ of NS matter and compare with the empirical $\Lambda$ deduced from the 
analysis of the GW data of GW170817 in Sect.~\ref{sec4}. The explicit treatment 
of the spin- and isospin variables in the CDM3Yn density dependent interaction
allows us to show explicitly the impacts by the spin symmetry energy $W$, nuclear 
symmetry energy $S$, and nuclear incompressibility $K$ to the EOS of nucleonic matter, 
the tidal deformability, mass and radius of NS.    

\section{Hartree-Fock calculation of the spin-polarized nuclear matter} 
\label{sec2}
The nonrelativistic Hartree-Fock (HF) approach \cite{Loan11} has been extended 
recently \cite{Tan20} to study the spin-polarized NM at zero temperature. In this case, 
NM is characterized by the neutron and proton number densities, $n_{\rm n}$ and 
$n_{\rm p}$, or equivalently by the total baryon number density $n_{\rm b}=n_{\rm n}+n_{\rm p}$ 
and neutron-proton asymmetry $\delta=(n_{\rm n}-n_{\rm p})/n_{\rm b}$. The spin 
polarization of baryons is treated explicitly for neutrons and protons by using the 
densities with baryon spin aligned up or down along the axis of magnetic field 
$\Delta_{\rm n,p}=(n_{\uparrow {\rm n,p}}-n_{\downarrow {\rm n,p}})/n_{\rm n,p}$. 
In general, the total HF energy density of NM is given by
\begin{align}
\mathcal{E}=\mathcal{E}_{\rm kin}+{\frac{1}{ 2}}\sum_{k \sigma \tau}
\sum_{k'\sigma '\tau '} [\langle{\bm k}\sigma \tau, {\bm k}' \sigma' \tau'
 |v_{\rm d}|{\bm k}\sigma\tau, {\bm k}' \sigma' \tau' \rangle   \nonumber\\
+ \langle{\bm k}\sigma \tau, {\bm k}'\sigma' \tau' |v_{\rm ex}|
{\bm k}'\sigma \tau, {\bm k}\sigma' \tau' \rangle], \label{eq1} 
\end{align}
where $|{\bm k}\sigma \tau\rangle$ are plane waves, and $v_{\rm d}$ and $v_{\rm ex}$ 
are the direct and exchange components of the (in-medium) density dependent 
NN interaction. Although the neutron and proton magnetic moments are of different 
strengths and of opposite signs, in the presence of strong magnetic field $|\Delta_n|$ 
and $|\Delta_p|$ should be of the same order. Like the previous HF study \cite{Tan20},
we also assume hereafter the baryon spin polarization 
$\Delta=\Delta_{\rm n}\approx -\Delta_{\rm p}$. 

We have used in the present work several versions of the density dependent CDM3Yn 
interaction which is based upon the (G-matrix) M3Y-Paris interaction \cite{Anan83}. 
These interactions were well tested in the earlier HF studies of NM \cite{Loan11,Tan20,Tan21} 
as well as the folding model analyses of nucleus-nucleus scattering \cite{Kho97,Kho00}. 
Explicitly, the CDM3Yn interaction is just the original M3Y-Paris interaction \cite{Anan83}
supplemented by an empirical density dependence \cite{Tan20,Kho97,Kho00}
\begin{align}
 v_{\rm d(ex)}(n_{\rm b}, r)=  F_{00}(n_{\rm b})v_{00}^{\rm d(ex)}(r) + 
 F_{10}(n_{\rm b}) v_{\rm 10}^{\rm d(ex)}(r)({\bm\sigma}\cdot {\bm\sigma}') \nonumber \\
 +F_{01}(n_{\rm b}) v_{01}^{\rm d(ex)}(r)({\bm\tau}\cdot{\bm\tau}')+F_{11}(n_{\rm b}) 
 v_{11}^{\rm d(ex)}(r)({\bm\sigma}\cdot {\bm\sigma}') ({\bm\tau}\cdot {\bm\tau}'). 
\label{eq2}
\end{align}
The radial dependence of the central interaction (\ref{eq2}) is determined from the 
spin (isospin) singlet and triplet components of the M3Y-Paris interaction \cite{Anan83}, 
and expressed terms of three Yukawa functions \cite{Kho96} as
\begin{equation}
 v^{\rm d(ex)}_{st}(r)=\sum_{\kappa=1}^3 Y^{\rm d(ex)}_{st}(\kappa)
\frac{\exp(-R_\kappa r)}{R_\kappa r}, \label{eq3}
\end{equation} 
The Yukawa strengths $Y^{\rm d(ex)}_{st}(\kappa)$ are given explicitly, e.g., 
in Table~I of Ref~\cite{Tan20}. If the spin polarization is neglected 
($\Delta=0$) then NM can be treated as \emph{spin-saturated}, and the 
$\sigma$-components of plane waves are averaged out in the HF calculation 
(\ref{eq1}). As a result, only the $st=00$ and $st=01$ terms of the central 
interaction (\ref{eq2}) are necessary for the determination of the energy density 
of NM. The situation becomes different when $\Delta\neq 0$, and the spin ($st=10$) 
and spin-isospin ($st=11$) dependent terms of the interaction (\ref{eq2}) need 
to be properly taken into account in the HF calculation. In this case, the total
HF energy density (\ref{eq1}) of the spin-polarized NM is obtained as
\begin{equation}
\mathcal{E}=\mathcal{E}_{\rm kin}+F_{00}(n_{\rm b})\mathcal{E}_{00} 
+ F_{10}(n_{\rm b})\mathcal{E}_{10} + F_{01}(n_{\rm b})\mathcal{E}_{01}
+ F_{11}(n_{\rm b})\mathcal{E}_{11} \label{eq4} 
\end{equation} 
The explicit expressions of $\mathcal{E}_{st}$ obtained with the density dependent 
CDM3Yn interaction are given in Ref.~\cite{Tan20}. We note that the \emph{isoscalar} 
density dependence $F_{00}(n_{\rm b})$ was first parametrized in Ref.~\cite{Kho97} 
to properly saturate symmetric NM at the density $n_0\approx 0.17$ fm$^{-3}$, while 
giving different values of the nuclear incompressibility $K$.  
The \emph{isovector} density dependence $F_{01}(n_{\rm b})$ was parametrized 
later to reproduce the microscopic Brueckner-Hartree-Fock (BHF) results of asymmetric 
NM, with the total strength fine tuned by the folding model description 
of the charge exchange $(p,n)$ reaction to the isobar analog states in finite 
nuclei \cite{Kho07,Kho14}. Because the spin polarization of baryons gives rise to 
the nonzero contribution from $\mathcal{E}_{10}$ and $\mathcal{E}_{11}$ to the total 
NM energy density (\ref{eq4}), the density dependencies $F_{10}(n_{\rm b})$ and 
$F_{11}(n_{\rm b})$ of the CDM3Y$n$ interaction (\ref{eq2}) need to be properly 
determined for the HF calculation of the spin-polarized NM. For the convenience 
in numerical calculations, the density dependent functional $F_{st}(n_{\rm b})$ 
of the CDM3Yn interaction (\ref{eq2}) has been assumed in the analytical form   
\begin{equation}
F_{st}(n_{\rm b}) = C_{st}\big[1+\alpha_{st}\exp(-\beta_{st}n_{\rm b}) 
+ \gamma_{st} n_{\rm b}\big].  \label{eq5}
\end{equation}
\begin{table} [bht]\vspace{-0.25cm}
\caption{Parameters of the density dependence (\ref{eq5}) of 4 versions of the CDM3Yn 
interaction used in the present HF calculation. $K$ is the nuclear incompressibility
\eqref{eq6} of symmetric NM determined at the saturation density $n_0\approx 
 0.17$ fm$^{-3}$.} \vspace{0cm}\label{t1}
\begin{center}
\begin{tabular}{ccccccc} \hline
  Interaction & $st$ & $C_{st}$ & $\alpha_{st}$ & $\beta_{st}$ & $\gamma_{st}$ & $K$\\
                & & & & (fm$^3$) & (fm$^3$) & (MeV)\\     \hline
    \multirow{4}{*}{\text{CDM3Y4}} & 00 & 0.3052 & 3.2998 & 2.3180 & -2.0 &\multirow{4}{*}{228}\\
                                   & 01 & 0.2129 & 6.3581 & 7.0584 & 5.6091 &\\
                                   & 10 & 0.1494 & 6.7055 & 2.5766 & 116.5455 &\\
                                   & 11 & 0.6830 & 0.6949 & 3.2104 & 1.0433 &\\
    \hline
    \multirow{4}{*}{\text{CDM3Y5}} & 00 & 0.2728 & 3.7367 & 1.8294 & -3.0 &\multirow{4}{*}{241}\\
                                   & 01 & 0.2204 & 6.6146 & 7.9910 & 6.0040 &\\
                                   & 10 & 0.1607 & 3.3867 & 2.8341 & 106.7274 &\\
                                   & 11 & 0.7016 & 0.6299 & 3.4552 & 1.0752 &\\
    \hline
    \multirow{4}{*}{\text{CDM3Y6}} & 00 & 0.2658 & 3.8033 & 1.4099 & -4.0 &\multirow{4}{*}{252}\\
                                   & 01 & 0.2313 & 6.6865 & 8.6775 & 6.0182 &\\
                                   & 10 & 0.1887 & -0.9998 & 3.1342 & 92.1075 &\\
                                   & 11 & 0.7259 & 0.5452 & 3.6416 & 1.0775 &\\
    \hline
    \multirow{4}{*}{\text{CDM3Y8}} & 00 & 0.2658 & 3.8033 & 1.4099 & -4.3 &\multirow{4}{*}{257}\\
                                   & 01 & 0.2643 & 6.3836 & 9.8950 & 5.4249 &\\
                                   & 10 & 0.2162 & -2.3396 & 3.3397 & 77.3144 &\\
                                   & 11 & 0.7573 & 0.4858 & 4.2011 & 1.0179 &\\ \hline
\end{tabular}
\end{center}
\end{table}
Such a procedure has been carried out for the CDM3Y8 version of the interaction in 
the HF calculation of the spin-polarized NM \cite{Tan20}, where the parameters 
of $F_{10}(n_{\rm b})$ and $F_{11}(n_{\rm b})$ were adjusted to reproduce the BHF 
results for the spin-polarized symmetric NM and neutron matter by Vida\~na 
{\it et al.} \cite{Vida16} using the Argonne V18 free NN potential added 
by the Urbana IX three-body force. In the present work, we apply the same procedure 
to 4 versions (CDM3Y4, CDM3Y5, CDM3Y6, and CDM3Y8) of the CDM3Yn interaction for
a comparative HF study. Note that the isovector density dependence $F_{01}(n_{\rm b})$ 
of these 4 interactions were fine tuned recently \cite{Tan21} to reach a good agreement 
of the nuclear symmetry energy given by the HF calculation with that given by the 
ab-initio calculations \cite{APR,MMC} at supra-saturation densities $n_{\rm b}>n_0$. 
The isoscalar density dependence $F_{00}(n_{\rm b})$ of these interactions has been 
kept unchanged as suggested in Ref.~\cite{Kho97}. All the parameters used in the present 
work are given explicitly in Table~\ref{t1}. 

\begin{figure}[bht]\vspace{-0.5cm}\hspace*{-0.5cm}
\includegraphics[angle=0,width=0.53\textwidth]{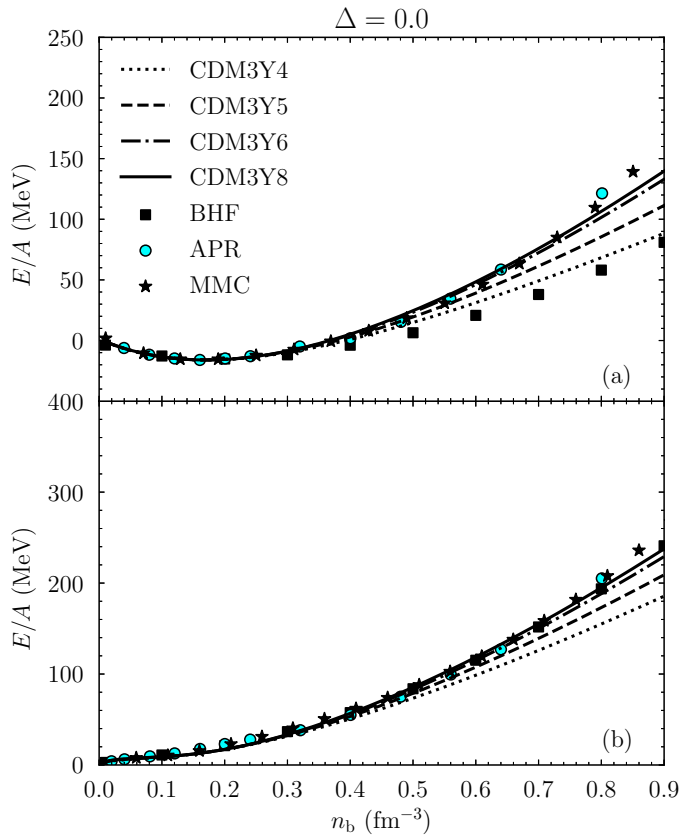}\vspace{-0.5cm}
 \caption{Energy per baryon of the spin-unpolarized symmetric NM (a) and 
pure neutron matter (b) given by the HF calculation (\ref{eq1}) using four 
versions of the CDM3Yn interaction, in comparison with the BHF results (squares) 
\cite{Vida16}. The circles and stars are results of the \textit{ab-initio} 
calculations by Akmal, Pandharipande, and Ravenhall (APR) \cite{APR} and microscopic 
Monte Carlo (MMC) calculation by Gandolfi {\it et al.} \cite{MMC}, respectively.} 
\label{f1}
\end{figure}
\begin{figure}[bht]\vspace{-0.3cm}\hspace*{-0.5cm}
\includegraphics[angle=0,width=0.53\textwidth]{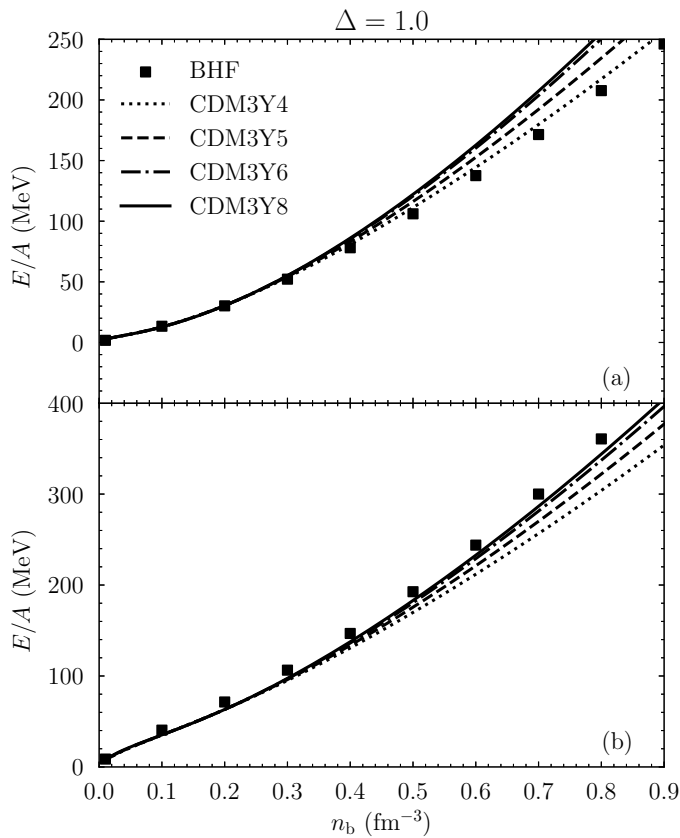}\vspace{-0.5cm}
 \caption{The same as Fig.~\ref{f1} but for the fully spin-polarized symmetric NM 
(a) and pure neutron matter (b), in comparison with the BHF results (squares) 
\cite{Vida16}.} \label{f2}
\end{figure}
\begin{figure}[bht]\vspace{-0.5cm}\hspace{-1cm}
\includegraphics[angle=0,width=0.55\textwidth]{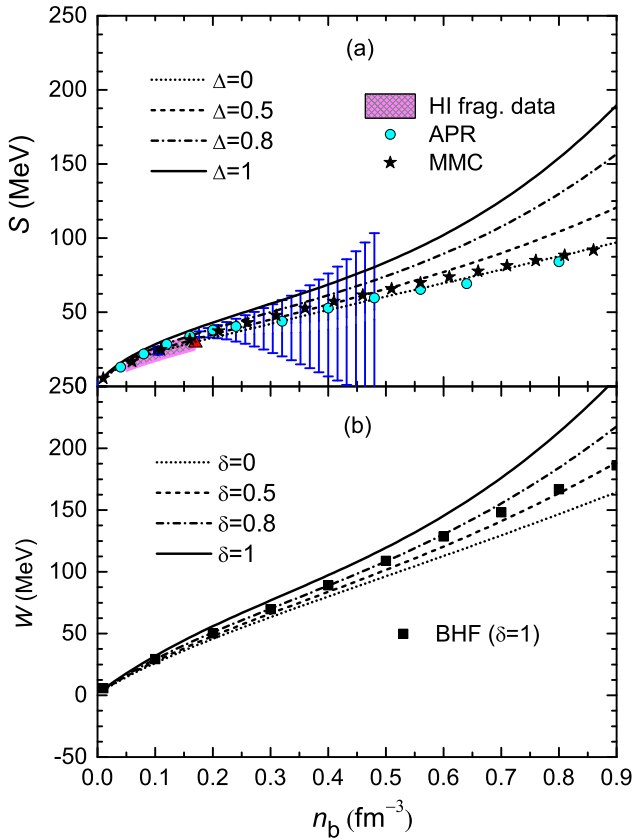}\vspace{-1cm}
 \caption{ (a) The HF results obtained with the CDM3Y8 interaction for the nuclear 
symmetry energy \eqref{eq7} at different spin polarizations of baryons, in comparison 
with the \textit{ab-initio} results \cite{APR,MMC} obtained with $\Delta=0$; the shaded 
region is the constraint by the HI fragmentation data \cite{Tsang11,Ono03}; the vertical 
bars are given at 90\% confidence level by the Bayesian analysis \cite{Xie19}. (b) 
The spin symmetry energy \eqref{eq8} given by the CDM3Y8 interaction at different 
isospin polarizations $\delta$, in comparison with the BHF results for the fully 
spin-polarized neutron matter \cite{Vida16}.} \label{f3}
\end{figure}

Dividing the total NM energy density \eqref{eq1} by the baryon number density $n_{\rm b}$
we obtain the NM energy per baryon $E/A$, which is shown for the spin-unpolarized and 
spin-polarized NM in Figs.~\ref{f1} and \ref{f2}, respectively. The energy of symmetric NM 
at high baryon densities correlates strongly with the nuclear incompressibility 
$K$ determined at the saturation density as 
\begin{equation}
 K=9n_{\rm b}^2 \frac{\partial^2}{\partial n_{\rm b}^2}
 \frac{E}{A}(\delta=0)\Bigr\vert_{n_{\rm b}\to n_0}
 =9\frac{\partial P(\delta=0)}{\partial n_{\rm b}}
 \Bigr\vert_{n_{\rm b}\to n_0}.  \label{eq6}
\end{equation}
Note that a slight difference of the HF results for the energy of symmetric NM 
at high baryon densities (upper panel of Fig.~\ref{f1}) is due to the different
values of the incompressibility $K$ obtained with 4 versions of the interaction 
(see Table~\ref{t1}). The $K$ value is strongly sensitive to the EOS of NM, and 
$K$ has been, therefore, a key research topic of numerous structure studies of 
nuclear monopole resonances (see, e.g., Ref.~\cite{Garg18} and references therein) 
as well as studies of the refractive light heavy-ion (HI) scattering \cite{Kho07r}. 
These researches have narrowed the empirical range to $K\approx 240\pm 20$ MeV. 
While the 4 density dependent versions of the CDM3Yn interaction (\ref{eq2}) give 
$K\approx 228 - 257$ MeV which are well within the empirical range, such a difference 
in $K$ values was shown to affect significantly the gravitational mass of NS obtained 
with the EOS given by these interactions \cite{Tan21}. In the present work we explore
this effect also for the EOS of the spin-polarized NM. One can see in Fig.~\ref{f2} that 
the full spin polarization of baryons ($\Delta=1$) substantially enhances the energy of both
the symmetric NM and pure neutron matter over the whole range of densities, and the energy 
required per baryon to change the spin-unpolarized NM into the fully spin-polarized NM
is the \emph{spin symmetry} energy $W$ \cite{Tan20}. Governed by the same SU(2) symmetry,
the spin symmetry energy behaves in about the same manner as the \emph{isospin symmetry} 
energy $S$, which is widely known as the nuclear symmetry energy $S$. Thus, the total 
energy of NM can be expressed alternatively as   
\begin{eqnarray}
\frac{E}{A}&=&\frac{E}{A}(n_{\rm b},\Delta,\delta=0)+ S(n_{\rm b},\Delta)\delta^2
+O(\delta^4)+... \label{eq7} \\
&=&\frac{E}{A}(n_{\rm b},\Delta=0,\delta)+ W(n_{\rm b},\delta)\Delta^2+O(\Delta^4)
+... \label{eq8} 
\end{eqnarray}
The contribution from both the higher-order terms $O(\delta^4)$ and $O(\Delta^4)$ 
were proven to be small and can be neglected in the well-known \emph{parabolic} 
approximation \cite{Tan20,Kho96}. In the present context, we find it illustrative 
to consider the pure neutron matter as the fully \emph{isospin} polarized NM ($\delta=1$), 
so that the nuclear symmetry energy $S(n_{\rm b})$ equals just the energy required per 
baryon to change the (isospin) symmetric NM into the fully isospin-polarized NM, i.e., 
the pure neutron matter. 
The (isospin) symmetry energy $S(n_{\rm b},\Delta)$, widely discussed in the literature 
as the {\it nuclear symmetry energy}, is the key characteristics of the EOS of NS matter 
and is, therefore, a longstanding goal of numerous nuclear physics studies (see, e.g., 
Refs.~\cite{BALi08,Hor14,Lat14}). However, these studies were done mainly for the 
spin-saturated NM, and describe, therefore, $S(n_{\rm b},\Delta=0)$.  
The HF results obtained with the CDM3Y8 interaction for the nuclear symmetry energy 
\eqref{eq7} at different spin polarizations of baryons $\Delta$ are compared in the upper
panel of Fig.~\ref{f3} with the \textit{ab-initio} results obtained at $\Delta=0$ 
\cite{APR,MMC} as well as the constraint implied by the analysis of the HI fragmentation 
data \cite{Tsang11,Ono03}. Using the parameters of $F_{01}(n_{\rm b})$ of the CDM3Yn
interaction fine tuned recently \cite{Tan21}, the HF results obtained with $\Delta=0$
agree nicely with those of the ab-initio calculations at high densities. The nuclear
symmetry energy increases significantly with the increasing spin polarization 
$\Delta$, but the $S(n_{\rm b},\Delta)$ values remain well within the empirical range 
inferred from a Bayesian analysis of the correlation of different EOS's of the np$e\mu$ 
matter with the GW170817 constraint on the radius $R_{1.4}$ of NS with mass $M=1.4~M_\odot$ 
\cite{Xie19} (the vertical bars in the upper panel of Fig.~\ref{f3}). It is natural to 
expect that this GW170817 constraint also has an imprint of the spin polarization 
of baryons in the two coalescing neutron stars.
The density dependence of the nuclear symmetry energy is widely investigated in terms 
of the symmetry coefficient $J$, slope $L$ and curvature $K_{\rm sym}$ of an expansion
of $S$ around the saturation density $n_0$ \cite{BALi08,Hor14,Lat14}
\begin{equation}
 S(n_{\rm b})=J+\frac{L}{3}\left(\frac{n_{\rm b}-n_0}{n_0}\right) +
 \frac{K_{\rm sym}}{18}\left(\frac{n_{\rm b}-n_0}{n_0}\right)^2+ ... \label{eq9}
\end{equation}
These quantities and the incompressibility $K$ of symmetric NM are the main characteristics 
of the EOS of NM. For the \emph{spin-unpolarized} asymmetric NM, the symmetry coeficient $J$ 
is well established to be around 30 MeV, but the $L$ and $K_{\rm sym}$ values remain much
less certain. A recent systematic survey by Li {\it et al.} \cite{BALi21} quotes 
$L\approx 57.7\pm 19$ MeV at the 68\% confidence level. 
The relativistic mean-field studies have suggested \cite{PREX} that the neutron skin 
thickness $R_{\rm skin}=R_{\rm n}-R_{\rm p}$ of the $^{208}$Pb nucleus is a stringent 
laboratory constraint on the slope $L$ of the symmetry energy $S(n_{\rm b})$. 
Given $R_{\rm skin}\approx 0.283\pm 0.071$ fm deduced recently from the 
measurement of the parity-violating asymmetry in the elastic scattering of polarized 
electrons from $^{208}$Pb by the PREX collaboration \cite{PREX2}, one obtains 
$L\approx 106\pm 37$ MeV using different relativistic energy density functionals 
\cite{PREX}. This $L$ value is significantly higher than that predicted by most 
of the mean-field calculations, and impacts strongly the calculated macroscopic 
properties of NS \cite{PREX}. 
 
At variance with the nuclear symmetry energy \eqref{eq7}, the spin symmetry energy \eqref{eq8} 
(see lower panel of Fig.~\ref{f3}) was much less studied, and we could compare the HF 
results for $W(n_{\rm b},\delta)$ only with the BHF result obtained for the fully 
spin-polarized neutron matter \cite{Vida16}. With the quadratic dependence on 
$\delta$ and $\Delta$ of the NM energy, the stiffness of the EOS of spin-polarized NM
increases significantly with the increasing polarization of spin ($\Delta$)
and isospin ($\delta$) of baryons. Such effect is well seen in the behavior of 
$S(n_{\rm b},\Delta)$ and $W(n_{\rm b},\delta)$ with the increasing $\Delta$ and 
$\delta$, respectively. Given such a correlation of the $S$ and $W$, it is obvious 
that the spin polarization of baryons should not be neglected in a mean-field study 
of NS matter. It is interesting that the density dependence of the spin symmetry energy 
can also be expressed in terms of the spin-symmetry coefficient 
$J_{\rm s}$, slope $L_{\rm s}$ and curvature $K_{\rm syms}$, in the same manner 
as the expansion \eqref{eq9}, 
\begin{equation}
 W(n_{\rm b})=J_{\rm s}+\frac{L_{\rm s}}{3}\left(\frac{n_{\rm b}-n_0}{n_0}\right) +
 \frac{K_{\rm syms}}{18}\left(\frac{n_{\rm b}-n_0}{n_0}\right)^2+ ... \label{eq10}
\end{equation}
$J_{\rm s}$, $L_{\rm s}$, and $K_{\rm syms}$ have {\it not} been 
investigated so far in different mean-field models of the EOS of NM. In the present 
work these quantities are obtained with 4 versions of the CDM3Yn interaction, and 
the those determined at $\delta=0$ are shown in Table~\ref{t2} together with
the symmetry coefficient, slope and curvature of the nuclear symmetry energy (\ref{eq9})
determined at $\Delta=0$. 
\begin{table} [bht]\vspace{-0.25cm}
\caption{The symmetry coefficient, slope and curvature of the nuclear symmetry energy 
(\ref{eq9}) for the spin-unpolarized asymmetric NM ($\Delta=0$), and those 
of the spin symmetry energy (\ref{eq10}) for the spin-polarized symmetric NM 
($\delta=0$) given by the present HF calculation using 4 versions of the CDM3Yn 
interaction.} \vspace{0cm}\label{t2}
\begin{center}
\begin{tabular}{ccccccc} \hline
  Interaction & $J$ & $L$ & $K_{\rm sym}$ & $J_{\rm s}$ & $L_{\rm s}$ & $K_{\rm syms}$ \\
            &  (MeV) & (MeV) & (MeV) & (MeV) & (MeV) & (MeV) \\     \hline
  {\text{CDM3Y4}} & 30.0 & 50.0 & -63.5 &  40.4 & 96.0 & -70.7 \\
  {\text{CDM3Y5}} & 30.0 & 50.0 & -52.1 &  40.3 & 96.5 & -67.7 \\
  {\text{CDM3Y6}} & 30.0 & 50.0 & -44.2 &  40.3 & 97.4 & -64.0 \\
  {\text{CDM3Y8}} & 29.9 & 49.5 & -31.4 &  40.1 & 96.7 & -63.3 \\ \hline
\end{tabular}
\end{center}
\end{table}   
Although the spin symmetry energy at the saturation density $W(n_0)=J_{\rm s}\approx 40$ 
MeV which is only about 10 MeV larger than $S(n_0)=J\approx 30$ MeV. The slope $L_{\rm s}$
of the spin symmetry energy is nearly twice that of the nuclear symmetry energy ($L$) 
and this makes the EOS of the spin-polarized NM much stiffer than that of the 
spin-unpolarized NM (see Figs.~\ref{f1} and \ref{f2}). 
\begin{figure}[bht]\vspace{-0.2cm}\hspace{-1cm}
\includegraphics[angle=0,width=0.55\textwidth]{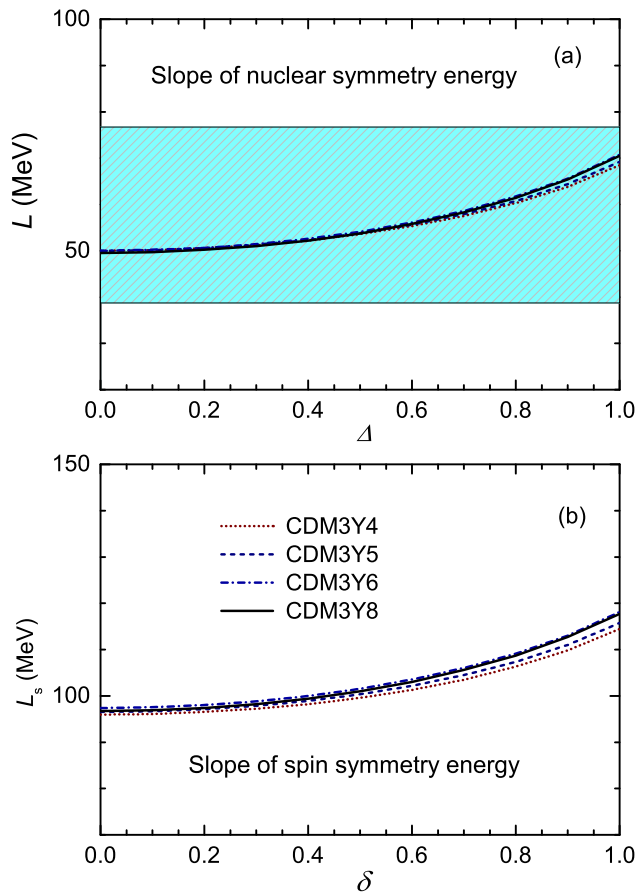}\vspace{-0.5cm}
 \caption{(a) The HF results obtained with 4 versions of the CDM3Yn interaction for: 
the slope $L$ of the nuclear symmetry energy \eqref{eq9} at different spin polarizations 
of baryons $\Delta$, in comparison with the empirical range suggested by Li {\it et al.} 
\cite{BALi21} at the 68\% confidence level (the shaded region). 
(b) The slope $L_{\rm s}$ of the spin symmetry energy \eqref{eq10} at different isospin 
asymmetries $\delta$.} \label{f4}
\end{figure}
In general, the parameters  $J,\ L$ and $K_{\rm sym}$ of the nuclear symmetry energy must 
depend on the spin polarization of baryons $\Delta$, and vice versa, $J_{\rm s},\ L_{\rm s}$ 
and $K_{\rm syms}$ also depend on $\delta$. Such a spin-isospin correlation of the two 
slope parameters is shown in Fig.~\ref{f4} and one can see about the same increasing 
trend of $L(\Delta)$ and $L_{\rm s}(\delta)$ with the increasing spin and isospin 
polarization, respectively. Over the whole range $0\lesssim\Delta\lesssim 1$, 
the obtained $L(\Delta)$ values remain well within the empirical range implied by the  
nuclear physics studies and astrophysical observations \cite{BALi21}, but are still 
below the lower limit of the $L$ value implied by the neutron skin of $^{208}$Pb 
measured in the PREX-2 experiment \cite{PREX,PREX2}.   

\begin{figure}[bht]\vspace{0cm}\hspace{-0.5cm}
\includegraphics[angle=0,width=0.5\textwidth]{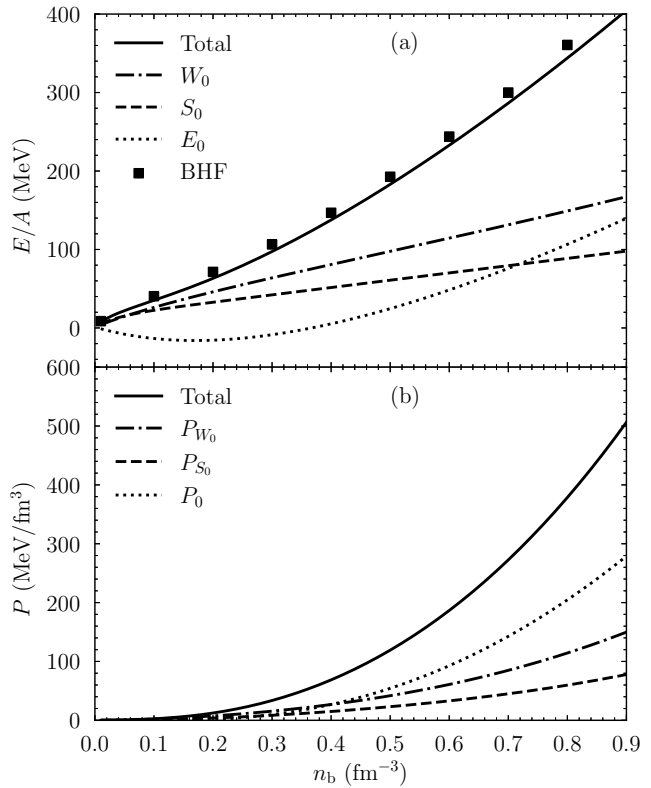}\vspace{-0.5cm}
\caption{(a) Energy per baryon \eqref{eq11} of the fully spin-polarized neutron matter 
given by the HF calculation using the CDM3Y8 interaction (solid line), and that given 
by the BHF calculation (squares) \cite{Vida16}. The energy per baryon $E_0$ of the 
spin-unpolarized symmetric NM, nuclear symmetry energy $S_0$ and spin symmetry energy 
$W_0$ are shown as dotted, dashed and dash-dotted lines, respectively. (b) The total
baryonic pressure of the fully spin-polarized neutron matter (solid line) in terms 
of the contributions obtained separately from $E_0$, $S_0$, and $W_0$.} \label{f5}
\end{figure}
As discussed above, the energy of the spin-polarized asymmetric NM depends on both 
the spin polarization of baryons $\Delta$ and the neutron-proton asymmetry 
(or the isospin polarization) $\delta$ in about the same manner. As a result, it turns 
out possible to expand $E/A$ simultaneously in the spin- and isospin polarizations, 
which enables the description of the energy of NM in terms of the nuclear symmetry 
\eqref{eq9} and spin symmetry energy \eqref{eq10} that depend on the baryon density only
\begin{eqnarray}
& &\frac{E}{A}=E_0(n_{\rm b})+S_0(n_{\rm b})\delta^2+W_0(n_{\rm b})\Delta^2+O(\delta^4)
+O(\Delta^4)+...,\nonumber\\
& & {\rm where}\ E_0(n_{\rm b})=\frac{E}{A}(n_{\rm b},\Delta=0,\delta=0),\nonumber\\
& & S_0(n_{\rm b})=S(n_{\rm b},\Delta=0),\ {\rm and}\ 
W_0(n_{\rm b})=W(n_{\rm b},\delta=0). \label{eq11} 
\end{eqnarray}
By comparing the full HF result for $E/A$ and that given by the parabolic 
approximation in the expansion \eqref{eq11}, we found that the contributions 
from the quartic and higher orders in $\delta$ and $\Delta$ are negligible  
over baryon densities up to $n_{\rm b}\approx 0.9$ fm$^{-3}$. This important 
feature shows a close similarity between the spin symmetry and isospin 
symmetry in the mean-field study of the spin-polarized NS matter. The explicit 
contributions of the spin-symmetry and isospin-symmetry energies to the EOS 
of the fully spin-polarized neutron matter ($\Delta=\delta=1$) are illustrated 
in Fig.~\ref{f5}, and one can see that $W_0(n_{\rm b})$ has about the same strength as that
of $S_0(n_{\rm b})$ at low baryon densities $n_{\rm b}\lesssim 0.1$ fm$^{-3}$. However, 
the spin symmetry energy becomes much stronger with the increasing $n_{\rm b}$, and 
$W_0$ is nearly double the nuclear symmetry energy $S_0$ at high baryon densities
which significantly stiffens the EOS of neutron matter. It can be seen in the lower 
panel of Fig.~\ref{f5} that the baryonic pressure of the fully spin-polarized neutron 
matter is also about twice that of the spin-unpolarized neutron matter. We note that 
the total energy per baryon \eqref{eq11} of the fully spin-polarized neutron matter 
given by the present HF calculation is quite close to that given by the microscopic 
BHF calculation \cite{Vida16} using the Argonne V18 potential supplemented 
by a realistic three-body force (see upper panel of Fig.~\ref{f5}). 
The nuclear symmetry energy $S_0(n_{\rm b})$ given by the HF calculation is also 
close to that given by the \textit{ab-initio} calculations \cite{APR,MMC} (see upper
panel of Fig.~\ref{f3}). Consequently, the expansion \eqref{eq11} should be 
of interest for the mean-field studies of the spin-polarized NS matter, where 
some estimate of the spin symmetry energy $W_0(n_{\rm b})$ can be done based on 
the expansion \eqref{eq10} using parameters given in Table~\ref{t2}. 

\section{EOS of the spin-polarized neutron star matter in $\beta$ equilibrium} 
\label{sec3}
The HF approach \eqref{eq1}-\eqref{eq4} describes NM that contains nucleons only. 
In fact, the NS matter contains significant lepton fraction in both the crust and uniform 
core, and a realistic EOS of NS matter should include the lepton contribution. For the 
inhomogeneous NS crust, we have adopted the EOS given by the nuclear energy density 
functional calculation \cite{Chamel18,Chamel19} using the BSk24 Skyrme functional, 
with atoms being fully ionized and electrons forming a degenerate Fermi gas. 
At the edge density $n_{\rm edge}\approx 0.06$ fm$^{-3}$, a weak first-order 
phase transition takes place between the NS crust and uniform core of NS. 
At baryon densities $n_{\rm b}\gtrsim n_{\rm edge}$ the core of NS is described as a 
homogeneous matter of neutrons, protons, electrons and negative muons ($\mu^-$ appear 
at $n_{\rm b}$ above the muon threshold density $\mu_e>m_\mu c^2\approx 105.6$ MeV). 
The total mass-energy density $\mathcal{E}$ of the spin-polarized np$e\mu$ matter is 
determined as
\begin{eqnarray}
& & \mathcal{E}(n_{\rm n},n_{\rm p},\Delta,n_e,n_\mu)=  
\mathcal{E}_{\rm HF}(n_{\rm n},n_{\rm p},\Delta)+n_{\rm n}m_{\rm n}c^2 \nonumber\\
& &\ +n_{\rm p}m_{\rm p}c^2+\mathcal{E}_e(n_e)+\mathcal{E}_\mu(n_\mu), \label{eq12} 
\end{eqnarray}
where $\mathcal{E}_{\rm HF}(n_{\rm n},n_{\rm p},\Delta)$ is the HF energy density 
(\ref{eq4}) of the spin-polarized baryonic matter, $\mathcal{E}_e$ and $\mathcal{E}_\mu$ 
are the energy densities of electrons and muons given by the relativistic Fermi gas 
model \cite{Shapiro}. In such a Fermi gas model, the spin polarization of leptons 
does not affect the total energy density $\mathcal{E}$, and the lepton number densities 
$n_e$ and $n_\mu$ are determined from the charge neutrality condition ($n_{\rm p}=n_e+n_\mu$), 
and the $\beta$-equilibrium of (neutrino-free) NS matter is sustained by the balance of the 
chemical potentials 
\begin{equation}
\mu_{\rm n} = \mu_{\rm p}+\mu_e\ \ \text{and}\ \ \mu_e=\mu_\mu,\ 
\text{where}\ \mu_j=\frac{\partial\mathcal{E}_j}{\partial n_j}. \label{eq13}
\end{equation} 
The fractions of the constituent particles in the spin-polarized np$e\mu$ matter 
are determined at the given baryon density $n_{\rm b}$ as $x_j=n_j/n_{\rm b}$. Below 
the muon threshold density ($\mu_e<m_\mu c^2\approx 105.6$ MeV) the charge neutrality 
condition leads to the following relation \cite{Loan11} 
\begin{equation}
 3\pi^2(\hbar c)^3n_{\rm b}x_{\rm p}-\hat\mu^3=0,\ 
 \hat\mu=\mu_{\rm n}-\mu_{\rm p}=2\frac{\partial}{\partial\delta}
 \left(\frac{\mathcal{E}_{\rm HF}}{n_{\rm b}}\right). \label{eq14}
\end{equation}
The proton fraction in the $\beta$-stable (spin-polarized) np$e\mu$ matter,
$x_{\rm p}(n_{\rm b},\Delta)$, can then be obtained from the solution of Eq.~(\ref{eq14}).
If we assume the parabolic approximation and neglect the contribution from
higher-order terms in (\ref{eq7}), then $x_{\rm p}(n_{\rm b},\Delta)$ is given by the
solution of the well-known equation 
\begin{equation}
 3\pi^2(\hbar c)^3n_{\rm b}x_{\rm p}-[4S(n_{\rm b},\Delta)(1-2x_{\rm p})]^3=0, 
 \label{eq15}
\end{equation}
which shows the direct link between the nuclear symmetry energy $S(n_{\rm b},\Delta)$
and the proton abundance in the NS matter. Above the muon threshold, 
$\mu_e > m_\mu c^2\approx 105.6$ MeV, it is energetically favorable for electrons 
to convert to negative muons, and the charge neutrality condition results on the 
following equation for $x_{\rm p}(n_{\rm b},\Delta)$ 
\begin{equation}
 3\pi^2(\hbar c)^3n_{\rm b}x_{\rm p}-\hat\mu^3-
 [\hat\mu^2-(m_\mu c^2)^2]^{3/2}\theta(\hat\mu-m_\mu c^2)=0, \label{eq16}
 \end{equation}
where $\theta(x)$ is the Heaviside step function. The proton fraction 
$x_{\rm p}(n_{\rm b},\Delta)$ is known to correlate with the NS cooling rate. 
In particular, the direct Urca (DU) process of NS cooling via neutrino emission 
is possible only if $x_{\rm p}$ is above the DU threshold $x_{\rm DU}$ \cite{Loan11}  
\begin{equation}
 x_{\rm DU}(n_{\rm b})=\frac{1}{1+\left[1+r_e^{1/3}(n_{\rm b})\right]^3}, 
 \label{eq17}
\end{equation}
where $r_e(n_{\rm b})=n_e/(n_e+n_\mu)$ is the leptonic electron fraction at the given
baryon number density. At low densities $r_e=1$, and $x_{\rm DU}\approx 11.1\%$,  
which is the muon-free threshold for the DU process. Since the lepton-baryon 
interaction is neglected in the present study, $x_{\rm DU}$ depends very weakly 
on the spin polarization of baryons $\Delta$. 
\begin{figure}[bht]\vspace{-0.5cm}\hspace*{-0.5cm}
\includegraphics[angle=0,width=0.52\textwidth]{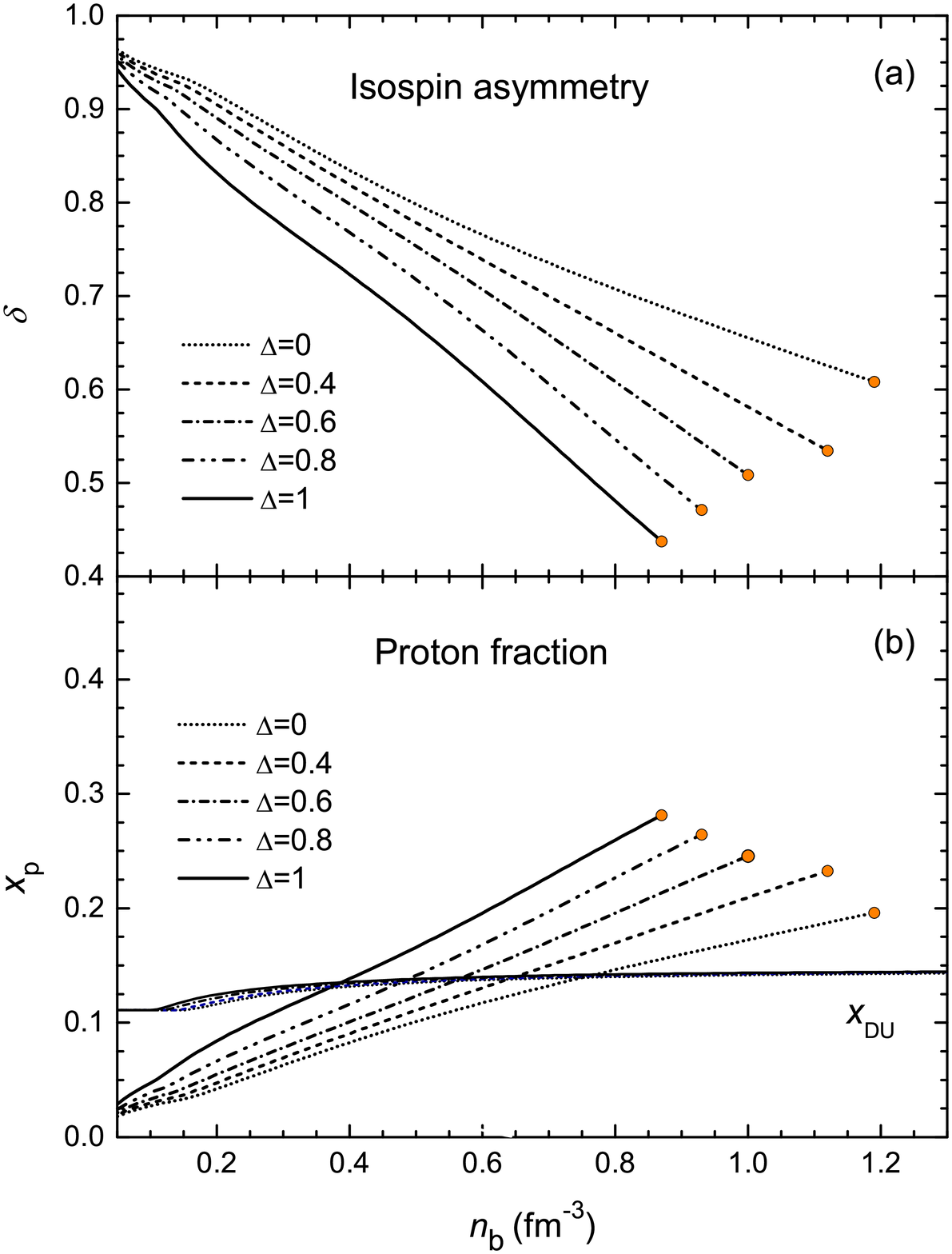}\vspace{-1cm}
 \caption{The isospin asymmetry $\delta$ (a) and proton fraction $x_{\rm p}$ (b) of 
the $\beta$-stable np$e\mu$ matter at different spin polarizations $\Delta$ of baryons
given by the HF calculation (\ref{eq12})-(\ref{eq15}) using the CDM3Y8 interaction. 
The circles are $\delta$ and $x_{\rm p}$ values obtained at the maximum central densities 
$n_{\rm c}$, and the thin lines are the corresponding DU thresholds (\ref{eq17}).} \label{f6}
\end{figure} 
Because the nuclear symmetry energy $S(n_{\rm b},\Delta)$ increases steadily with 
the increasing spin polarization of baryons $\Delta$ (see Fig.~\ref{f3}), from 
Eqs.~\eqref{eq14}-\eqref{eq16} one can expect the same trend for the proton fraction 
$x_{\rm p}$ of the spin-polarized $\beta$-stable np$e\mu$ matter. As shown in the 
lower panel of Fig.~\ref{f6}, $x_{\rm p}$ increases significantly with the increasing 
spin polarization of baryons, and it exceeds the DU threshold at densities 
$n_{\rm b}\gtrsim 2n_0$ for the fully spin-polarized NS matter with $\Delta=1$. 
It is also natural that the neutron-proton asymmetry $\delta$ decreases with the 
increasing $x_{\rm p}$ as shown in the upper panel of Fig.~\ref{f6}. The charge 
neutrality implies also an increasing electron fraction with the increasing $x_{\rm p}$, 
up to 20\%$-$30\% when $\Delta\lesssim 1$. Such a high electron fraction was 
found in the blue kilonova ejecta following GW170817 \cite{Abb17a,Abb17b,Evan17}, and 
suggested by Metzger \textit{et al.} \cite{Metzger18} to be of the magnetar origin. 

\begin{figure}[bht]\vspace{0cm}\hspace*{-1.50cm}
\includegraphics[angle=0,width=0.63\textwidth]{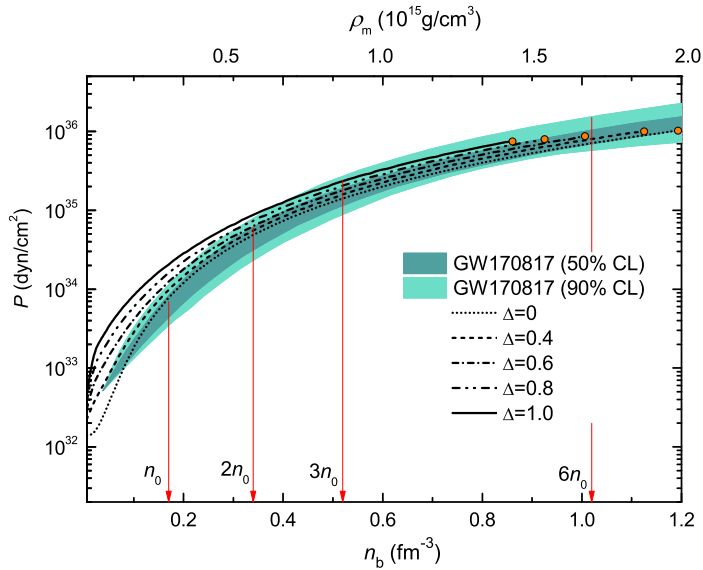}\vspace{0cm}
 \caption{The total pressure (\ref{eq18}) inside the core of NS at different spin 
polarizations $\Delta$ given by the HF calculation using the CDM3Y8 interaction, over 
the range of the mass-energy ($\rho_{\rm m}$) and baryon-number ($n_{\rm b}$) densities. 
The dark and light shaded regions are the empirical constraints given by the ``spectral" 
EOS inferred from the Bayesian analysis of the GW170817 data at the 50\% and 90\% confidence 
levels, respectively \cite{Abb18}. The circles are $P(n_{\rm c},\Delta)$ values 
at the corresponding maximum central densities $n_{\rm c}$.} \label{f7}
\end{figure} 
The EOS of the spin-polarized np$e\mu$ matter in $\beta$ equilibrium is determined 
entirely by the mass-energy density 
$\rho_{\rm m}(n_{\rm b},\Delta)=\mathcal{E}(n_{\rm b},\Delta)/c^2$ and 
the total pressure  
\begin{equation}
P(n_{\rm b},\Delta)=n_{\rm b}^2{\frac{\partial}{{\partial n_{\rm b}}}}
\left[\frac{\mathcal{E}_{\rm HF}(n_{\rm b},\Delta)}{n_{\rm b}}\right]+P_e+P_\mu. 
\label{eq18}
\end{equation}
We show in Fig.~\ref{f7} the total pressure (\ref{eq18}) of the spin-polarized NS matter 
$P(n_{\rm b},\Delta)$ obtained with different $\Delta$ values from the HF calculation 
using the CDM3Y8 interaction over baryon densities up to above $6n_0$, in comparison 
with the empirical pressure given by the ``spectral" EOS inferred from the Bayesian 
analysis of the GW signals of GW170817 at the 50\% and 90\% confidence levels 
\cite{Abb18}. One can see the substantial impact by the spin symmetry energy \eqref{eq8} 
with the increasing $\Delta$, which stiffens the EOS and compresses nucleonic matter 
inside the NS core to the higher pressure over the whole range of densities. It is
noticeable that $P(n_{\rm b},\Delta)$ obtained with $0.8\lesssim\Delta\lesssim 1$ 
overestimates the empirical constraint at the baryon densities within the range 
$0.05n_0\lesssim n_{\rm b}\lesssim 2n_0$.
 
\section{Tidal deformability, mass and radius of neutron star}
\label{sec4}
The interesting effect inferred from the GW170817 observation is the tidal deformation 
of NS induced by the strong gravitational field which enhances the GW emission and 
accelerates the decay of the quasicircular inspiral \cite{Abb18}.   
We recall briefly the tidal deformability of a static spherical star being exposed to the 
gravitational field created by the attraction of the companion star in a binary system
\cite{Hind08,Hind10}. At the range close enough, this star is tidally deformed 
and gains a nonzero quadrupole moment $Q_{ij}$ that is directly proportional to the  
strength $E_{ij}$ of the gravitational field 
\begin{equation}
 Q_{ij} = -\lambda E_{ij}. \label{eq19}
\end{equation}
$\lambda$ characterizes the star response to the gravitational field and is dubbed 
as the \emph{tidal deformability} or quadrupole polarizability of star. In the General
Relativity, $\lambda$ is related to the $l=2$ tidal Love number $k_2$ \cite{Hind08} as
\begin{equation}
 \lambda = \frac{2}{3G}k_2 R^5, \label{eq20}
\end{equation}
where $R$ and $G$ are the star radius and gravitational constant, respectively. 
It is convenient to consider the (dimensionless) tidal deformability parameter  
$\Lambda$ \cite{Abb18} expressed in terms of the compactness $C$ of star with mass 
$M$ and radius $R$ as 
\begin{equation}
 \Lambda=\frac{2}{3}k_2C^{-5}, \ {\rm with}\ C=\frac{GM}{Rc^2}. \label{eq21}
\end{equation}
Using the linearized Einstein equation, the Love number $k_2$ can be expressed 
in terms of the nonzero metric perturbation of the stress-energy tensor $H(r)$ 
and its radial derivative $H'(r)$ \cite{Hind08,Hind10}, which are determined 
from the solution of a differential equation that is integrated together with 
the Tolman-Oppenheimer-Volkoff equations. More details on this computation can
be found, e.g., in Ref.~\cite{Tan21}.   
\begin{figure}[bht]\vspace{0cm}\hspace*{-1.0cm}
\includegraphics[width=0.62\textwidth]{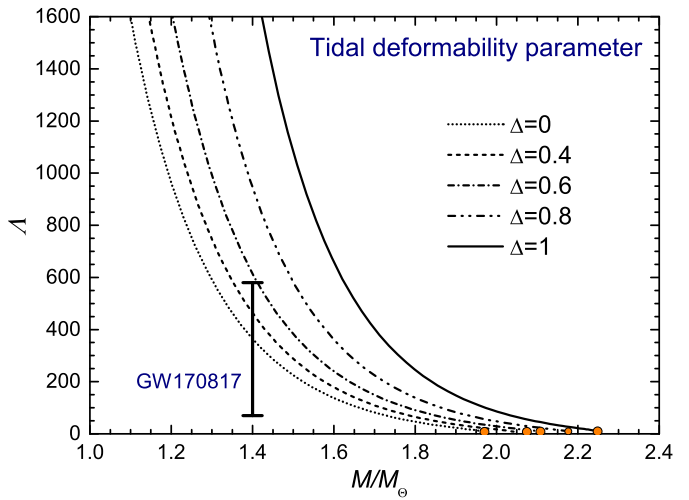}\vspace*{-0.5cm}
 \caption{The tidal deformability parameter (\ref{eq21}) given by the EOS of the
spin-polarized NS matter obtained with different $\Delta$ values from the HF 
calculation using the CDM3Y8 interaction. The vertical bar is the empirical $\Lambda$ 
value for NS with $M=1.4~M_\odot$ inferred from the Bayesian analysis of the GW170817 
data at the 90\% confidence level \cite{Abb18}, and the circles are $\Lambda$ values 
obtained at the corresponding maximum central densities $n_{\rm c}$.} \label{f8}
\end{figure}
\begin{figure}[bht]\vspace{-1.3cm}\hspace*{-1cm}
\includegraphics[width=0.62\textwidth]{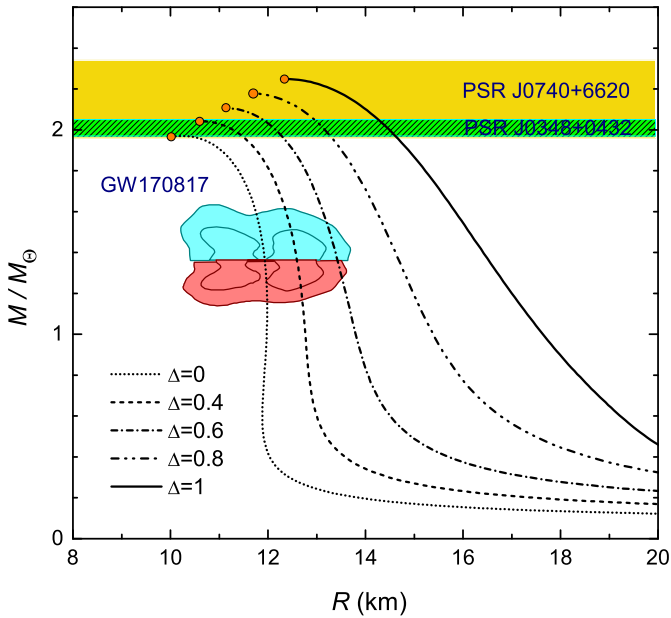}\vspace*{-0.5cm}
\caption{The gravitational mass of NS versus its radius given by the EOS of the
spin-polarized NS matter obtained with different $\Delta$ values from the HF 
calculation using the CDM3Y8 interaction. The colored contours are the GW170817 
constraint for NS with mass $M=1.4~M_\odot$ \cite{Abb18}, and the circles 
are $M$-$R$ values calculated at the corresponding maximum central densities 
$n_{\rm c}$. The shaded areas are the observed masses of the second PSR J$0348+0432$ 
\cite{Ant13} and millisecond PSR J$0740+6620$ \cite{Cro20}.} \label{f9} 
\end{figure}

The tidal deformability and gravitational mass-radius of NS given by the EOS of the
spin-polarized NS matter obtained with different $\Delta$ values from the HF 
calculation using the CDM3Y8 interaction are shown in Figs.~\ref{f8} and \ref{f9}, 
respectively. One can see that the heavier the NS the smaller its tidal deformability,
and the impact by the spin symmetry energy to $\Lambda$ is very well revealed in
Fig.~\ref{f8}, where only the $\Lambda$ values given by the EOS's of the partially
spin-polarized NS matter with $\Delta\lesssim 0.6$ are inside the empirical range 
implied by the GW170817 data at the 90\% confidence level for NS with $M=1.4~M_\odot$
\cite{Abb18}. Like the tidal deformability, the NS mass and its radius comply 
with the GW170817 constraint for NS with mass $M=1.4~M_\odot$ \cite{Abb18} (the colored 
contours in Fig.~\ref{f9}) when the EOS's of the partially spin-polarized NS matter 
with $\Delta\lesssim 0.6$ are used for the input of the TOV equations. 
The stiffening of the EOS of the spin-polarized NS matter by the spin symmetry energy 
shown in Fig.~\ref{f6} is well reflected in the calculated $M$-$R$ values, and the 
obtained maximum masses $M$ of NS (solid circles in Fig.~\ref{f9}) span the whole 
empirical range of the masses deduced for the second PSR J$0348+0432$ \cite{Ant13} 
and millisecond PSR J$0740+6620$ \cite{Cro20}, the heaviest neutron stars observed 
so far. The results shown in Figs.~\ref{f8} and \ref{f9} confirm that the GW170817 
constraint excludes the full spin polarization of baryons ($\Delta=0.8\sim 1$) inside 
the core of NS, as pointed out earlier in Refs.~\cite{Tan20,Tews20}. 

\subsection*{Density dependence of the spin polarization}
The above results were obtained with the uniform spin polarization of baryons which 
is independent of the baryon density $n_{\rm b}$. However, nucleons inside the inner 
core of NS are known to be fully degenerate and occupy all possible quantum states 
allowed by the Pauli principle \cite{Glen00}. Such a full degeneracy exhausts all 
spin orientations of baryons and a spin polarization (or asymmetric spin orientation 
of baryons with $\Delta>0$) is unlikely inside the inner core of NS. As a result, 
$\Delta$ inside the core of NS must be dependent on the baryon density. 
Moreover, the distribution of magnetic field inside magnetar was shown to be quite 
complex \cite{Fuji14}, and the spin polarization of baryons is expected to decrease 
gradually to $\Delta\approx 0$ in the central region of magnetar where the intensity 
of magnetic field is diminishing to zero \cite{Fuji14}. 
Although it is beyond the scope of the present mean-field approach to properly 
calculate the density profile $\Delta(n_{\rm b})$ of the spin polarization 
of baryons in magnetized NS, we try to explore this effect by assuming a realistic 
scenario for the density dependence of $\Delta$ based on the magnetic-field 
distribution in magnetar obtained by Fujisawa and Kisaka using the Green's function 
relaxation method \cite{Fuji14}. 

\begin{figure}[bht]\vspace{0.0cm}\hspace*{0.0cm}
\includegraphics[width=0.5\textwidth]{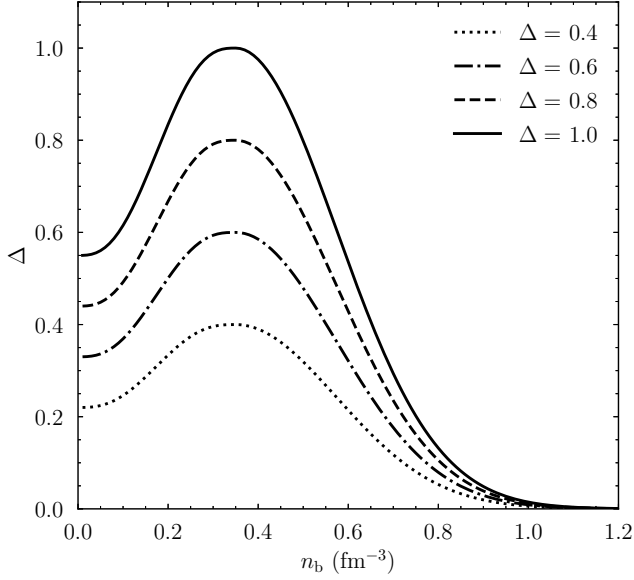}\vspace*{-0.5cm}
\caption{Density dependence of the spin polarization $\Delta$ of baryons inside
the core of NS that mimics the distribution of magnetic field in NS obtained 
by Fujisawa and Kisaka using the Green function relaxation method \cite{Fuji14}, 
with different maximum $\Delta$ values.} 
\label{f10} 
\end{figure}

Several simple scenarios of $\Delta(n_{\rm b})$ having its maximum at the surface and
decreasing gradually to zero towards the center of NS were considered in Ref.~\cite{Tan20}, 
and it was shown that up to 60\% of baryons might have their spins polarized during the 
NS merger GW170817. A more elaborate density dependence of $\Delta(n_{\rm b})$ is 
suggested in the present work, based on the spatial distribution of the magnetic flux 
$\Psi$ derived in Ref.~\cite{Fuji14}. Namely, the radial distribution of $\Psi$ from 
the star center to the surface (see Fig.~5 in Ref.~\cite{Fuji14}) has first been 
translated into the density profile of this quantity, then it is reasonable to assume
the density profile of $\Delta(n_{\rm b})$ to have the same shape as that 
of $\Psi(n_{\rm b})$. In such a scenario, the spin polarization of baryons $\Delta$ 
and the intensity of magnetic field reach their maximum values at the same baryon density 
(around $2n_0$). The asymmetric spin orientation $\Delta$ is expected to decrease gradually 
to zero in the inner core of NS where baryons are believed to be fully degenerate 
\cite{Glen00}. To explore the impact of the spin symmetry energy, we have probed 
different maximum values of $\Delta(n_{\rm b})$ at its peak as shown in Fig.~\ref{f10}. 
 
\begin{figure}[bht]\vspace{-1.3cm}\hspace*{-1cm}
\includegraphics[width=0.62\textwidth]{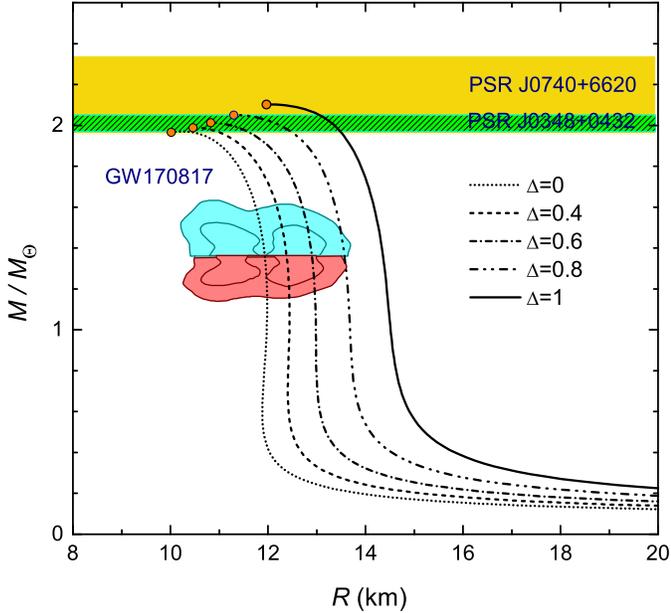}\vspace*{-0.5cm}
\caption{The same as Fig.~\ref{f9} but obtained with the density-dependent 
spin polarization $\Delta(n_{\rm b})$ of different strengths shown in 
Fig.~\ref{f10}.} \label{f11} 
\end{figure}
We show in Fig.~\ref{f11} the gravitational mass and radius of NS given by the EOS 
of the $\beta$-stable spin-polarized NS matter obtained with different (density dependent) 
spin polarizations of baryons $\Delta(n_{\rm b})$ shown in Fig.~\ref{f10}. Following 
the trend of $\Delta(n_{\rm b})$, the strength of the spin symmetry energy \eqref{eq10}
is also diminishing to zero in the inner core of NS, and the impact of the spin polarization 
of baryons on the maximum mass of magnetar becomes weaker compared to that shown 
in Fig.~\ref{f9}. Based on the suggested scenario for $\Delta(n_{\rm b})$, we found 
that the gravitational mass and radius of NS given by the EOS of the spin-polarized 
NS matter with $\Delta\lesssim 0.8$ are well within the empirical range implied for NS 
with $M=1.4~M_\odot$ \cite{Abb18}. In general, NM becomes less compressible \cite{Tan20}
when the spin polarization of baryons is nonzero, and NS expands its size with 
the maximum mass $M_{\rm max}$ and radius $R_{\rm max}$ becoming larger with the 
increasing $\Delta$ as shown in Fig.~\ref{f9}. With the damping of the $\Delta$ strength 
in the inner core of NS shown in Fig.~\ref{f10}, the impact of the spin symmetry energy 
at high densities becomes less significant, and the enhancement of the NS mass and 
radius (see Fig.~\ref{f11}) with the increasing maximum $\Delta$ value is not as drastic 
as shown in Fig~\ref{f9}. At variance with the spin symmetry energy, the impact of the 
nuclear symmetry energy to the EOS of NS matter remains still strong at large baryon 
densities where the $\delta$ value sustained by the $\beta$ equilibrium is up to 
0.6 (see upper panel of Fig.~\ref{f6}). Thus, the nuclear symmetry energy $S(n_{\rm b})$ 
at high baryon densities in the inner core of NS, where the spin symmetry energy 
$W(n_{\rm b})$ decreases quickly to zero, is the most important input for the EOS 
of the $\beta$-stable NS matter. 
\begin{figure}[bht]\vspace{-1.0cm}\hspace*{-0.5cm}
\includegraphics[angle=0,width=0.55\textwidth]{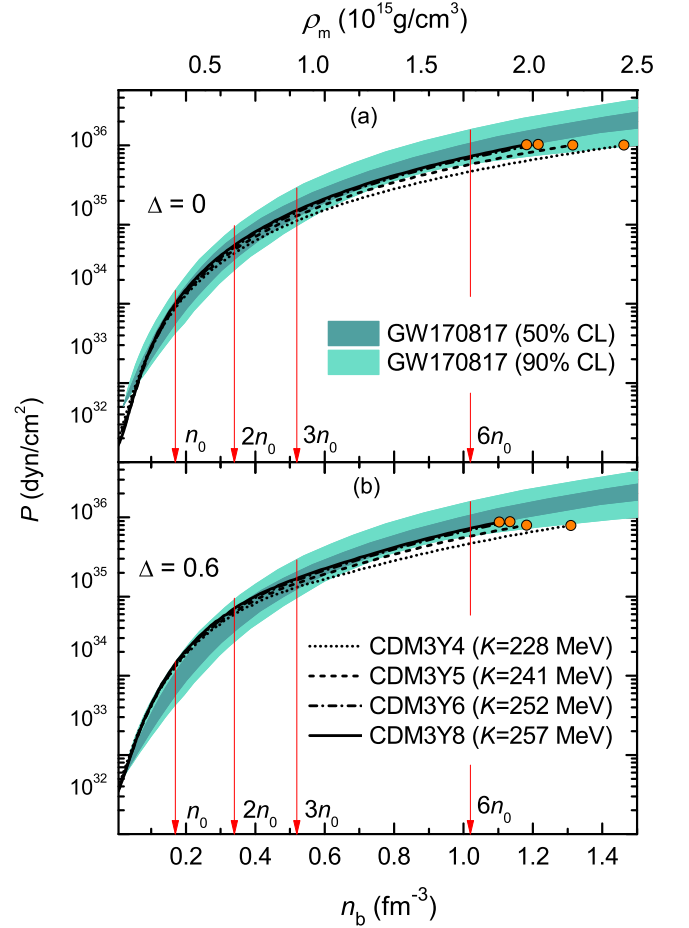}\vspace{-1.0cm}
 \caption{The same as Fig.~\ref{f7} but obtained with 4 versions of the CDM3Yn 
density dependent interaction (\ref{eq2})-(\ref{eq5}) that are associated with
4 different values of the nuclear incompressibility $K$. (a) - the results
obtained for the spin unpolarized NS matter; (b) - the results obtained for 
the partially spin polarized NS matter with the maximum $\Delta=0.6$ of the
spin polarization strength shown in Fig.~\ref{f10}.} \label{f12}
\end{figure} 
\begin{figure}[bht]\vspace{-0.5cm}\hspace*{-1.0cm}
\includegraphics[angle=0,width=0.6\textwidth]{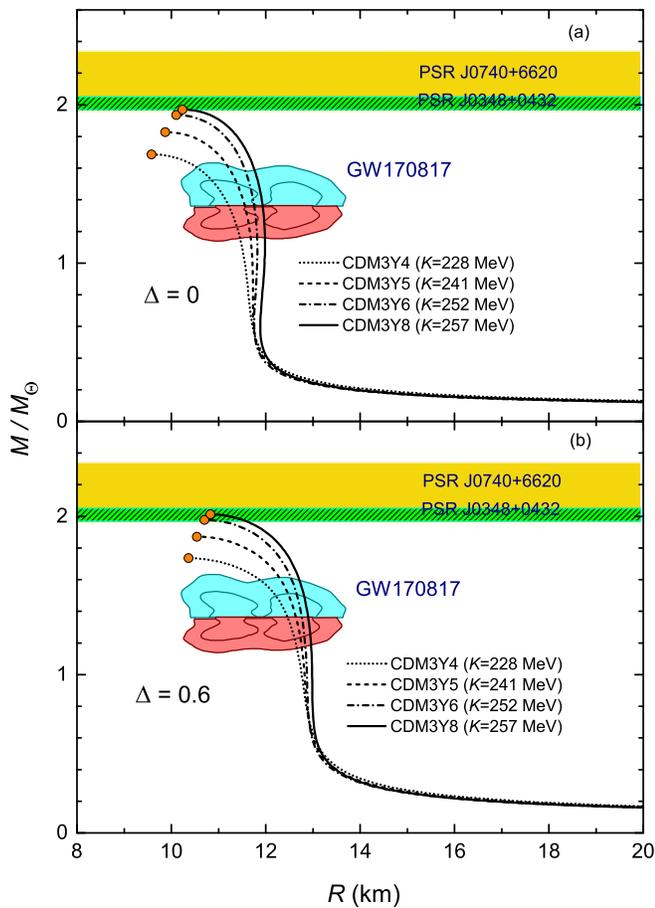}\vspace{-1.0cm}
 \caption{The same as Fig.~\ref{f9} but obtained with 4 versions of the CDM3Yn 
density dependent interaction (\ref{eq2})-(\ref{eq5}) that are associated with
4 different values of the nuclear incompressibility $K$. (a) - the results
obtained for the spin-unpolarized NS matter; (b) - the results obtained for 
the partially spin-polarized NS matter with the maximum $\Delta=0.6$ of the
spin-polarized fraction of baryons shown in Fig.~\ref{f10}.} \label{f13}
\end{figure}

\subsection*{Impact of the nuclear incompressibility}
Although the impact of the nuclear incompressibility on the EOS of NM is well 
known as shown in Fig.~\ref{f1}, it is of interest to explore explicitly this 
effect on the calculated macroscopic properties of NS. The total pressure 
(\ref{eq17}) inside the uniform core of NS obtained with 4 versions of the CDM3Yn 
density dependent interaction (\ref{eq2})-(\ref{eq5}), associated with 4 different 
values of the nuclear incompressibility $K$, is shown in Fig.~\ref{f12}.
One can see that the difference caused by different $K$ values is significant at high
baryon densities $n_{\rm b}>2n_0$ for both the spin-unpolarized and spin-polarized
cores of NS. Such a difference in the pressure results on quite different maximum 
masses of NS, and the results shown in Fig.~\ref{f13} suggest that a slightly
stiffer EOS of NS matter associated with $K\approx 250-260$ MeV not only complies with 
the GW170817 constraints but also gives the maximum mass of NS close to $2M_\odot$,
at the lower mass limit of the heaviest pulsars observed so far \cite{Ant13,Cro20}. 
Although the partial spin polarization of baryons expands the size of NS and increases 
the radius $R_{1/4}$ up to about 1 km, the difference in the NS maximum masses up 
to $0.3~M_\odot$ shown in the upper and lower panels of Fig.~\ref{f13} is mainly 
due to the difference in the $K$ values. In connection with these results, we note 
that Annala {\it et al.} have suggested recently \cite{Annala20} that the NS matter 
in the interior of massive NS with $M\approx 2~M_\odot$ might contain a quark-matter 
core that contributes up to $0.25~M_\odot$ to the total mass of NS. In any case, 
the present mean-field results are complementary to those of a joint analysis 
of the NICER and LIGO/Virgo data \cite{Raa20} that prefers a stiff EOS associated 
with the observed masses of the heaviest pulsars.    

\section*{Summary}  
Equation of states of the spin-polarized NM is studied within the HF formalism 
using the realistic CDM3Yn density dependent interaction. Given the nonzero
spin polarization or asymmetric spin orientation of baryons, the spin- and spin-isospin 
dependent terms of the HF energy density give rise to the spin symmetry energy $W$
which behaves in a manner similar to that of the isospin- or nuclear symmetry 
energy $S$. The parabolic approximation is shown to be valid also for the spin symmetry 
energy, so that the (repulsive) contribution from the spin symmetry energy to the total 
NM energy is directly proportional to $\Delta^2$. The EOS of NM becomes much stiffer 
with the increasing spin polarization of baryons, with the pressure given by the spin
symmetry energy at high baryon densities being much larger than that 
given by the nuclear symmetry energy.   

Like the nuclear symmetry energy \eqref{eq9}, the density dependence of the spin 
symmetry energy can also be expressed \eqref{eq10} in terms of three quantities: 
the symmetry coefficient $J_{\rm s}$, slope $L_{\rm s}$, and curvature $K_{\rm syms}$. 
A close correlation of these charateristics with those of the nuclear symmetry 
energy $S$ has been discussed, in particular, a very similar behavior of the slope 
parameters $L$ and $L_{\rm s}$ of $S(n_{\rm b})$ and $W(n_{\rm b})$, respectively. 
The slope $L$ of the nuclear symmetry energy depends strongly on the spin polarization 
of baryons $\Delta$, and the slope $L_{\rm s}$ of the spin symmetry energy depends 
on the isospin polarization $\delta$ in exactly the same manner. The $L$ values 
obtained at $0\leqslant\Delta\leqslant 1$ comply well with the constraint inferred from 
the nuclear physics studies and astrophysical observations \cite{BALi21}, but remain 
below the lower limit of $L$ implied by the neutron skin of $^{208}$Pb measured 
in the PREX-2 experiment \cite{PREX,PREX2}. 

With the EOS of the $\beta$-stable np$e\mu$ matter of NS obtained at different 
spin polarization of baryons, we found that the proton fraction $x_{\rm p}$ increases 
strongly with the increasing $\Delta$, which should result on the larger probability 
of the direct Urca process in the cooling of the magnetar. The charge neutrality 
then implies an increasing electron fraction with the increasing $x_{\rm p}$ 
that might reach up to around 30\%. 

The stiffening of the EOS of the NS matter with the increasing spin polarization  
of baryons affects significantly the calculated tidal deformability as well as the 
gravitational mass and radius of NS. This effect of the spin symmetry energy is 
very strong if we assume a uniform (density independent) spin polarization of baryons 
in both the outer and inner cores of NS. In such a scenario, the GW170817 constraint 
for the tidal deformability, mass and radius of NS with $M=1.4~M_\odot$ \cite{Abb18} 
excludes the spin polarization of baryons inside the core of NS with 
$0.6\lesssim\Delta\lesssim 1$. 

A more realistic scenario of $\Delta(n_{\rm b})$ is further suggested, based on 
the distribution of magnetic field inside magnetar obtained in Ref.~\cite{Fuji14}
and the full degeneracy of baryons in the inner core of NS \cite{Glen00},
where $\Delta$ reaches its maximum in the outer core and decreases quickly 
to zero in the inner core of NS. By subjecting the mean-field results obtained
at different spin polarizations of baryons in this scenario to the GW170817 constraint, 
we found that up to 80\% of baryons in the outer core of NS might have their spins 
polarized during the NS merger.  

The impact by the nuclear incompressibility $K$ on the macroscopic properties 
of NS is shown clearly for both the spin saturated ($\Delta=0$) and spin polarized 
($\Delta\neq 0$) NS matter. While the $M$-$R$ results given by 4 versions of the 
CDM3Yn interaction comply well with the GW170817 constraint deduced for NS with 
mass $M=1.4~M_\odot$ \cite{Abb18}, only the EOS associated with $K\approx 250-260$
MeV gives the maximum mass of NS close to $2M_\odot$, at the lower mass limit of the 
second PSR J$0348+0432$ \cite{Ant13} and millisecond PSR J$0740+6620$ \cite{Cro20}, 
the heaviest neutron stars observed so far.  

\section*{Acknowledgement}
The present research was supported, in part, by the National Foundation for
Science and Technology Development of Vietnam 
(NAFOSTED Project No. 103.04-2021.74). 
 
\bibliographystyle{apsrev4-2}
\bibliography{refK}
\end{document}